\documentstyle [12pt,a4]{article}
\newcommand{\vs}[1]{\rule[- #1 mm]{0mm}{#1 mm}}

\newcommand{\vs}[1]{\rule[- #1 mm]{0mm}{#1 mm}}
\begin{document}
\begin{titlepage}
\newpage
\setcounter{page}{0}

\vs{8}

\begin{center}
{\Large {\bf The Infinite Symmetry and Interplay\\[.5cm]
Between Integer and Fractional\\[.5cm]
Quantum Hall Effect}}\\[2cm]

{\large M.Eliashvili$^{1}$} \\[ .5cm]
{\it Laboratoire de Physique Th\'eorique}
{\small E}N{\large S}{\Large L}{\large A}P{\small P}^\ddag \\
{\it Chemin de Bellevue, B.P. 110, F-74941 Annecy-le-Vieux, Cedex, France.}

\end{center}

\vs{10}

\centerline{\bf Abstract}

\indent

We have revised the procedure of the explicit construction of the
$W_{1+\infty}$ for the QHE. It is shown, that the algebra representations
for the fractional and integer filling fractions are related by the
non-unitary similarity transformation. This transformation corresponds
to the introduction of the complex Chern-Simons gauge potentials, in
terms of which the second quantized form of FQHE can be developed.

\vs{8}

\rightline{{\small E}N{\large S}{\Large L}{\large A}P{\small P}-A-462/94}
\rightline{HEP-TH/9404070}
\rightline{March 1994}
\vspace*{\fill}

$\;$
\hrulefill\ $\; \; \; \; \; \; \; \; \; \; \; \; \; \; \; \; \; \; \; \;\;$
\hspace*{3.5cm}\\
\noindent
{\footnotesize $\;\ddag$ URA 14-36 du CNRS, assosi\'e  \`a l'E.N.S. de
Lyon, et au L.A.P.P. (IN2P3-CNRS) d'Annecy-le-Vieux
 \\

{\footnotesize $\;^1$ On leave of absence from Tbilisi Mathematical Institute,
  Tbilisi 380093, Georgia}

\end{titlepage}
%\end{document}

%\documentstyle[12pt,fa4]{article}
%\begin{document}

\begin{itemize}
\item

The current understanding of the quantum Hall effect is essentially based
on the Laughlin's picture of the incompressible two dimensional quantum
fluid, which exhibits an energy gap\cite{Laugh}.

In the recent series of papers Cappelli, Trugenberger and Zemba \cite{Cap1}-
\cite{Cap2}
 have
related the notion of incompressibility to the infinite symmetry, which
on the classical level is represented by the group of area preserving
 diffeomorphisms (see also \cite{Iso}).

 As a outcome two dimensional quantum fluid can be characterized
by the unitary irreducible highest weight representations of the $W_{1+\infty}$
algebra \cite{Cap3}.

The derivation of this basic conclusion is straightforward for the IQHE, when
liquid is formed by the noninteracting planar electrons in the lowest Landau
level $(\nu=1)$. For the clarity and to fix the notations we'll reproduce
some essential points.

The Hamiltonian of N electrons in the orthogonal uniform magnetic field
 $B=\varepsilon _{\alpha \beta}\partial_\alpha A^{\beta}$ ($ {\alpha},{\beta}
=1,2$)
is given by

$$
H=\frac {1}{2m}\sum_{k=1}^{N}[{\bf p}_k - e{\bf }A({\bf r}_k)]^2 =
\frac {1}{m}\sum_{k=1}^{N}(a_k a^+_k +a^+_k a_k),
$$

where

$$
a=\frac {i}{2}[(p_x +ip_y) -e(A_x+iA_y)] \equiv i[p - \frac {e}{2}A]
$$

$$
a^+=-\frac {i}{2}[(p_x-ip_y)-e(A_x-iA_y)] \equiv -i[\bar p -\frac {e}{2}\bar A]
$$

In the appropriately chosen system of units $(c=\hbar=m=1,e=2,B=1)$, and
symmetric gauge ${\bf {A}}=\frac {B}{2}(-y,x)$ \/ \rm
 the quantum-mechanical Hamiltonian and angular
momentum
 can be written in terms of harmonic oscillator operators:
$$
\hat {H}=\sum_{k=1}^{N}[a^+_ka_k + a_ka^+_k]
$$

$$
\hat {J}=\sum_{k=1}^{N}[b^+_kb_k - a^+_ka_k]
$$
In the complex notations $z=x+iy, \partial =\frac{1}{2}(\partial_x -i\partial
_y)$, these operators are given by
$$
a_k=\frac{z_k}{2} +\bar \partial_k \hspace{2cm}
 a^+_k=\frac{{\bar z_k}}{2} -\partial_k
$$

$$
b_k=\frac{\bar z_k}{2} +\partial_k \hspace{2cm}  b^+_k=\frac{z_k}{2} -
\bar\partial_k
$$

They satisfy the usual commutation relations:
$$
[a_k,a^{+}_l]=[b_k,b^+_l]=\delta_{kl}
$$
The $W_{1+\infty}$ is generated by the operators
$$
v^i_n=-\sum_{k=1}^N(b^+_k)^{n+i}(b_k)^i,\hspace{2cm} i\geq 0, \hspace{1cm}
n+i\geq 0
$$
 These operators commute with the Hamiltonian and satisfy the commutation
relations
\begin{equation}
[ v^i_n,v^j_m]=(jn-im)v^{i+j-1}_{n+m}+\cdot \cdot \cdot
\end{equation}
where multidots correspond to the quantum deformations \cite{Cap1}.

The $\nu=1$ ground state is given by the Laughlin's wave function \cite{Laugh}:
\begin{equation}
\Psi_1 (z_1,...,z_N)=\prod_{1\leq k<l\leq N}(z_k-z_l)e^{-1/2 \sum_k|z_k|^2}
\end {equation}

$$
\hat{H}\Psi_1=E_0\Psi_1= N \Psi_1
$$
$$
\hat {J}\Psi_1=\frac {N(N-1)}{N}\Psi_1
$$
The action of generators $v^i_n$ on the wave function (2) can be easily
 calculated (especially if on uses
second quantization formalism). The basic results are as follows
\cite{Cap1},\cite{Cap3}:

\begin{equation}
\begin{array}{l}
a)\hspace{15pt}v^i_n\Psi_1=0\vspace{5mm}
\hspace{3,5cm}for\hspace{1cm}-i\leq n<0,i\geq1  \\

\vspace {5mm}

b)\hspace{15pt} v^i_0\Psi_1=const\cdot\Psi_1 \\

\vspace {5mm}

c)\hspace{15pt} v^i_n\Psi_1=\Phi^i_n(z_1,...,z_N)\cdot\Psi_1  \hspace{2cm} for
  \hspace{1cm} n\geq 0,i\geq1
\end{array}
\end{equation}

Here $\Phi^i_n(z_1,...,z_N)$ is some symmetric polynomial.

The equality $a)$ in (3) is \it the highest weight condition,\/ \rm which is
the
mathematical transcription of the incompressibility. $b)$ and $c)$ characterize
the excitation spectrum.~

\item
The situation is drastically changed in the case of fractional fillings.
Now the ground state (for $\nu=\frac{1}{2p+1}, p$-integer) is given by
the Laughlin's wave function
\begin{equation}
\Psi_p(z_1,...,z_n)=\prod_{1\leq k<l\leq
 N}(z_k-z_l)^{2p+1}e^{-1/2\sum_k|z_k|^2}
\end{equation}
and is believed to describe the incompressible state of \it interacting\/ \rm
electrons (see e.g \cite{Stone}).

Now if one wants to construct the algebraic classification of quantum fluid,
the ground state (4) must be subjected to the action of the symmetry
generators. In order to carry out these calculations, in the recent paper
Flohr and Varnhagen \cite {Flo}
 have changed the definition of operators $b_k$, introducing an
interaction term:
\begin{equation}
b_k\Longrightarrow B_k=b_k-2p\sum_{l\neq k}\frac{1}{z_k-z_l}
\end{equation}
Note, that $b^+_k$ is not changed
\begin{equation}
b^+_k\Longrightarrow B^+_k=b^+_k
\end{equation}

Now the basic commutators contain the delta-function type terms:
$$
[B_k,B^+_l]=\delta_{kl}[1+2p\pi\sum_{k\neq j}\delta(z_k-z_j)]+
(\delta_{kl}-1)2p\pi\delta(z_k-z_l)
$$

The infinite symmetry is generated by the operators
$$
V^i_n=-\sum_{k=1}^N(B^+_k)^{n+i}(B_k)^i,
$$

which satisfy the same algebra as $v^i_n$ in (1), up to the terms involving
delta-functions. These terms can be ignored, because the wave functions vanish
as $z_k \rightarrow z_l$. As a result, it can be shown that $V^i_n$ acts
on $\Psi_p$ as on the highest weight state.

The operators $B_k$ and $B^+_k$  are not Hermitian conjugate. This will be
 improved, if one introduces the new integration
measure in the configuration space, i.e
$$
dz_1 \cdot \cdot \cdot dz_N\Longrightarrow dz_1 \cdot \cdot \cdot dz_N \mu
(z,\bar{z}),
$$
where
$$
\mu(z,\bar{z})=\prod_{k< l}|z_k-z_l|^{-4p}
$$
and simultaneously changes the definition of operators $a_k$ and $a^+_k$ in the
following way:
$$
a_k\Longrightarrow A_k=a_k
 $$
$$
a^+_k \Longrightarrow A^+_k=a^+_k +2p\sum_{k\neql}\frac{1}{z_k-z_l_}
$$

(for the details and further consideration see \cite{Flo}).~

\item
Remark, that the newly introduced operators $B_k$ act on the ground state
of interacting electrons $\Psi_p$ in a way analogous to the action of
 $b_k$'s on the $\Psi_1$:
$$
b_k\Psi_1=\sum_{k\neq j}\frac{1}{z_k-z_j}\Psi_1
$$
$$
B_k\Psi_p=\sum_{k\neq j}\frac{1}{z_k-z_j}\Psi_p
$$

It seems  that this circumstance had initiated the \it Ansatz\/-\rm
 type substitutions (5-6), which in turn leads to the introduction
of the measure $\mu(z,\bar z)$ and operators $A_k$ and $A^+_k$.

Now we can make a simple statement, which perhaps clarifies the meaning of
 this procedure: \it the wave functions and algebra generating operators
for the fractional $(\nu=\frac{1}{2p+1})$ and integer $(\nu=1)$ filling
factors are related by the following similarity transformation:\/  \rm
\begin{equation}
\Psi_p(z_1,...,z_N)=S_p(z_1,...,z_N)\Psi_1(z_1,...,z_N)
\end{equation}
\begin{equation}
\hat O_p=S_p(z_1,...,z_N){\hat {O}}_1 S^{-1}_p(z_1,...,z_N),
\end{equation}
where
\begin{equation}
S_p(z_1,...,z_N)=\prod_{k<l}(z_k-z_l)^{2p}
\end{equation}

(7) is evident, and (8) can be easily verified by the direct calculations,
 letting ${\large{ \hat O}}_1=\{b_k,b^+_k
a_k,a^+_k, v^i_n\}$ and ${\hat O_p}=\{B_k,B^+_k,A_k,A^+_k,V^i_n\}$
respectively.

Transformation (9) becomes singular as $z_k\rightarrow z_l$, but it acts
in the space of functions, which vanish in that limit. What seems to be
more important, is that it is not an unitary
transformation
$$
S^{\dagger}_pS_p=\prod_{k<l}|z_k-z_l|^{4p}=\mu(z,\bar z)^{-1}
$$

Evidently $\Psi_p$ is an eigenfunction of the transformed Hamiltonian

$$
\hat{H}_p=\hat {H} +4p\sum_{i\neq k} \frac {1}{z_i-z_k}a_i+
2p\pi\sum_{i\neq k}\delta(z_i-z_k),
$$

which must be considersd as a Hamiltonian of interacting electron system.

\em The transformations (7-9) interconnect the ground state vectors and
spectum generating quantum operators corresponding to two different
 physical phenomena: IQHE can be understood using a picture of
noninteracting electrons, while FQHE is essentially manifestation of
 interelectron interactions.
 On the other hand one can say, that from the point of view of
algebraic classification in the sense of \/ \rm\cite {Cap3}, \em the
IQHE and FQHE are non-unitary equivalent realizations of one and
the same underling symmetry. \/ \rm

Interesting to note, that non-unitary similarity transformations
have been recently considered by Ellis, Mavromatos and Nanopoulos
\cite{Ellis} in the context of quantum gravity, where they are
related to the temporal evolution between unstable quantum backgrounds,
indicating a deep conection between string quantum gravity and
incompressible Hall fluid.~

\item

Note, that the classical analogue of the non-unitary transfomations (8) is
given by the \it {non-canonical, complex}\/ \rm transformations
of the phase space variables
$z_k=x_k+iy_k$ and conjugated momenta $\bar p_k=\frac{1}{2}(p_{kx}-ip_{ky})}$:

$$
z_k\rightarrow z_k,\hspace {2cm}\bar p_k\rightarrow \bar p_k+i2p\sum_{l\neq
k}\frac{1}
{z_k-z_l}
$$

$$
\bar {z}_k\rightarrow \bar {z}_k, \hspace {2cm}  p_k\rightarrow p_k
$$

The substitutions $p\rightarrow p-\frac {e}{2}f$,
 $\bar p\rightarrow \bar p -\frac {e}{2}\bar f$
  can be inerpreted as an introduction of a \em {complex,
nonlocal vector potentials}\/\rm

\begin{equation}
f_k({\bf r}_1,...,{\bf r}_N)=f_{kx} +if_{ky}=0
\end{equation}
\begin{equation}
\bar{f}_k({\bf r}_1,...,{\bf r}_N)=f_{kx} - if_{ky}=-i2p\sum_{l\neq k}
\frac{1}{z_k-z_l}
\end{equation}

which depend on the positions of all N particles.

Magnetic field associated to these potentials, which acts on the k-th particle,
is given by the curl:

\begin{equation}
{\cal B}_k=i({\bar{\partial}_k} {\bar f}_k -{\partial}_kf_k)=
2p \pi \sum_{k\neq l}\delta(z_k-z_l)
\end{equation}

i.e. each particle sees the $N-1$ others as vortices carring a  $2p$ elementary
flux quanta. In the chosen system of units flux quantum $\phi_0=\pi$,
and the density of Landau states $n_B=\frac{eB}{2\pi}=1/\pi$. Hence,
the filling fraction

$$
\nu=\frac{N}{(\Phi/\phi_0)}=\frac{1}{2p+1},
$$

where the total flux $\Phi=\pi N B +2p\pi N =\pi N(2p+1)$.

Using the mean-field arguments, on can say that electrons move
in the average magnetic field $2p+1$, in accordance with the Jain's
hierarchical construction \cite{Jain}. However  the additional
magnetic field ${\cal B}=2p$ is generated now by the \em complex gauge
potentials,
\/\rm in contrast to the composite fermion approach \cite{Jain},
 where magnetic fluxes
attached to the point particles are produced by the real singular
vector potentials.

Note, that potential (11) has form
typical for the statistical interaction with a parameter $\theta =4\pi p$ (see
e.g.\cite{Lerda}).
So, if one tries to incorporate these potentials into
 the framework of the
Chern-Simons theories, one has to manage with the \em {complex gauge fields and
transformations}.\/ \rm

Here we'll give a brief account of this consideration.  Introduce the
 particle density at the point ${\bf r}$:

\begin{equation}
\varrho({\bf r})=\sum_{l=1}^N\delta({\bf r}-{\bf r}_l)
\end{equation}

and vector potentials  satisfying the equations:

\begin{equation}
\varepsilon_{\alpha \beta}\partial_{\alpha}f^\beta({\bf r})=2p\pi \varrho({\bf
r})
\end{equation}

and

\begin{equation}
\partial_\alpha f^\alpha({\bf r})=-i2p\pi\varrho ({\bf r})
\end{equation}

The solutions to (14)-(15) can be easily found:

\begin{equation}
f^\alpha({\bf r})=-2p\pi(\varepsilon_{\alpha \beta}+i\delta_{\alpha \beta})
\partial_{\beta}\int d{\bf r'}G({\bf r}-{\bf r'})\varrho ({\bf r'})
\end {equation}

Here $G({\bf r})=\frac{1}{2\pi}\ln r$ is a Green function.

Substituting into (16) the particle density  (13) and letting ${\bf r}=
{\bf r}_k$, we
immediately obtain (10)-(11).

The same time (14) is a field equation for the Chern-Simons Lagrangian

 \begin{equation}
{\cal L}=i\psi^\dagger(\partial_0+ieA_0+ief_0)\psi -
\frac {1}{2} |(\partial_\alpha + ieA_\alpha + ief_\alpha )\psi |^2 -
\frac {e^2}{8\pi p} {\varepsilon}^{\mu \nu \lambda}f_\mu \partial_\nu f_\lambda
\end{equation}

in the \em complex gauge \/ \rm (15). (Obviously $\varrho ({\bf r})$  must be
substituded by the current $J^0(x)=\psi^\dagger(x)\psi (x)$).
 The resulting gauge theory must lead
to the field-theoretic description of FQHE.

As a first step in this direction consider the problem in the mean-field
approximation. For this purpose substitute the local density $\varrho ({\bf
r})$ by its average value $\bar \varrho =n_B=1/\pi$.
 The corresponding one-particle Hamiltonian will be given by

\begin{equation}
H=\frac{1}{2}[{\bf p}-e{\bf A}-e{\bf f}]^2
\end{equation}
where the gauge potentials satisfy the following conditions:

$$
\varepsilon_{\alpha \beta}\partial_\alpha A^{\beta}=1, \hspace {3cm}
\partial_{\alpha}A^\alpha =0
$$

$$
\varepsilon_{\alpha \beta}\partial_\alpha f^{\beta}=2p, \hspace {3cm}
\partial_{\alpha}f^\alpha =-i2p
$$
Obviously, this potentials are given by

$$
{\bf A}=\frac {1}{2}(-y,x), \hspace {3cm} {\bf f}={\bf a}-{\bf \nabla}\lambda
$$
where
$$
{\bf a}=(-py,px), \hspace{3cm} \lambda=\frac {i}{2}p|z|^2
$$

The potential ${\bf f}$ can be considered as a gauge equivalent to ${\bf a}$,
with a purely imaginary gauge function $\lambda$. Consequently
the egenfunctions $\Phi$ of the Hamiltonian $H$ can be obtained as a
non-unitary "gauge" transformation

\begin{equation}
\Phi=e^{-ie\lambda}\Psi=e^{p|z|^2}\Psi
\end{equation}
 of eigenfunctions $\Psi$ of the Hamiltonian

$$
H_0=\frac{1}{2}[{\bf p}-e({\bf A}+{\bf a})]^2,
$$
which describes a particle moving in the magnetic field $B_{tot}=2p+1$.
Recalling, that
we are interested in the lowest Landau level, this functions are given by

\begin{equation}
\Psi_j(z.\bar z)\sim z^je^{-(2p+1)|z|^2/2}
\end{equation}
and correspond to the motion with a momentum $j$.

 Due to (19), the corresponding eigenfunctions of the initial Hamiltonian (18)

\begin{equation}
\Phi_j \sim z^j e^{-|z|^2/2}
\end{equation}
 coincide with the wave functions of particle moving in the
magnetic field $B=1$.

 Consider the $N$ particle state with a total momentum
$$
J=(2p+1)\frac {N(N-1)}{2},
$$
or equivalently the filling fraction
$$
\nu=\frac{1}{2p+1}.
$$

This state can  be constructed from the antisymmetrized product of the one-
particle wave functions (21) under the condition $\sum_{k=1}^Nj_k = J$.
Laughlin function (4) corresponds to the decomposition
of $J$ into into the sum of integers $j_k=(2p+1)(k-1)$, and in the second
quantized form is given by
$$
|\Psi_p\rangle \sim f^+_0f^+_{2p+1}\cdot \cdot \cdot f^+_{(2p+1)j}\cdot \cdot
\cdot
f^+_{(2p+1)(N-1)}|0 \rangle ,
$$
where $f^+_j$ creates a fermion state (21) with a momentum $j$.

Consequently the Laughlin's wave function $\Psi_p$ can be viewed
as a mean-field approximation for the $N$-particle state in the Chern-Simons
theory with a complex gauge potentials.

Similar consideration relates the $\nu=m$ states and operators to the
representation of $W_{1+\infty}$ at $\nu=\frac{m}{2mp+1}$. The corresponding
similarity transfomation will be given by
\begin{equation}
S_{p,m}=\prod_{I<J} \prod_{i<j}(z^I_i-z^J_j)^{K_{I,J}}\prod_I
\prod_{i<j}(z^I_i-z^I_j)^{K_{I,I}-1}
\end{equation}
where the $m\times m$ matrix $K_{IJ}$ is defined by \cite{Fro}

$$
  K=\left|
\begin{array}{cccc}
2p+1 & 2p & ... & 2p \\
2p   &2p+1 &... & 2p \\
 .   &  .  & .  & . \\
2p  &...   & 2p &2p+1
\end{array}
\right|
$$

Concluding, we can say the following: \sf According to \cite{Cap3}
  the quantum states
 of the
incompressible fluid can be exhaustively classified by the unitary irreducible
highest weight representations of the algebra $W_{1+\infty}$. Applying to the
representation at $\nu=$integer the similarity transformation (8) or
(18) one
 automatically (at least in principle)
 obtains the corresponding classification for the fractional values
of filling factor. This transformation seems to be equivalent to the
introduction of the complex C-S gauge potentials in terms of
which the field-theoretic formulation of FQHE can be developed.\/ \rm

\end{itemize}

\vspace{2cm}
{\Large {\bf Acknowledgements}}

\vspace{0.5cm}

I'm particularly thankful to P.Sorba for bringing the problems considered above
to my attention, many helpful discussions and constant encouragement.
It is a pleasure to thank N. Mavromatos for a useful discussion and D. Arnaudon
for interest.
The author wishes to express his gratitude to
ENSLAPP for the
 hospitality.

  \end{document}